\documentclass[aps,prb,twocolumn]{revtex4}
\usepackage{graphicx}

\begin{document}

\title{Spectral engineering with multiple quantum well structures}
\author{L. I. Deych}
\author{M. V. Erementchouk}
\author{A. A. Lisyansky}

\affiliation{Department of Physics, Queens College, City
University of New York, Flushing, New York 11367}

\begin{abstract}
It is shown that it is possible to significantly modify optical
spectra of Bragg multiple quantum well structures by introducing
wells with different exciton energies. The reflection spectrum of
the resulting structures is characterized by high contrast and
tuning possibilities.
\end{abstract}

\maketitle

\paragraph*{Introduction}

Band-gap engineering is aimed at creating materials with
pre-determined electrical properties. From the point of view of
optical or optoelectronic applications, it is also important to be
able to design materials with pre-determined optical spectra. This
requires a development of spectral engineering to control the
interaction between light and matter. The realization that such a
possibility exists led to the development of the field of photonic
crystals.\cite{Yablonovitch,John} Additional opportunities in
controlling the light-matter interaction arise in photonic
structures made of materials with internal resonances lying in the
spectral region of the photonic band
structure.\cite{DeychPRE1999,SivachenkoPRA2001,HuangPRL2003} Light
propagates through such materials in the form of polaritons, in
which electromagnetic waves are coupled with internal excitations
of the materials. By changing properties of the material
excitations, one can manipulate the properties of light as well.
Combining polaritonic effects with photonic crystal effects, one
obtains a greater flexibility in designing optical properties and
an opportunity to tune them after the growth. An interesting
one-dimensional example of such systems is given by Bragg multiple
quantum wells (BMQW).\cite{IvchenkoMQW} In these structures, the
wavelength of the quantum well (QW) exciton radiation, $\lambda$,
matches the period of the multiple quantum well structure, $d$:
$\lambda=2d$. As a result, the radiative coupling between quantum
wells causes a very significant modification of exciton radiative
properties, which are effectively controlled by geometrical
parameters of the structure. Such structures, therefore, are good
candidates for spectral engineering. In
Ref.~\onlinecite{DefectMQW} it was found that by replacing a
single base structural element of BMQW structure with an element
with different properties (a defect), one can significantly alter
optical spectra of these structures. It was shown that upon
introducing different types of the defect, a great variety of
spectral types could be
created.\cite{DefectMQW,DeychNanotechnology} However,
Ref.~\onlinecite{DefectMQW} dealt with ideal structures and it was
not clear if these effects can be reproduced in realistic
structures suffering from homogeneous and inhomogeneous
broadenings, and whose lengths are limited by technological
capabilities. The goal of this paper is to show that realistic
BMQW structures with defects (DBMQW) can be designed to exhibit
reflection spectra with sharp features characterized by high
contrast even in the presence of relatively large values of the
broadenings.  We will also show that the spectra of these
structures can be tuned after their growth with the help of the
quantum confined Stark effect (QCSE).\cite{QCSE} This makes DBMQW
structures a potential candidate for spectral engineering with
applications for tunable switching and modulating devices.

\paragraph*{Reflection spectra of DBMQW structures}
We consider a structure consisting of $N=2m+1$ QW-barrier layers.
The layers are identical except for one in the the middle, where
the quantum well has a different exciton frequency. While our
calculations are of a rather general nature, we will have in mind
a $GaAs/AlGaAs$ system as an example. In this case, such a defect
can be produced either by changing the concentration of
$Al$\cite{ALdependence_book,ALdependence} in the barriers
surrounding the central well, or the width of the well
itself\cite{widthdependence} during growth. While both these
methods will also affect the optical width of the defect layers,
this effect is negligibly small for the systems under
consideration, and we will assume that the exciton frequency is
the only parameter differentiating the defect well from the
others.

The reflection spectra are calculated using the transfer matrix
approach. The inhomogeneous broadening of the QW excitons is taken
into account within the framework of the effective medium
approximationá\cite{KavokinGeneral} which was shown to describe
the main  contribution to the reflection
coefficientþ\cite{ErementchoukPRB2003} Within this approach the
exciton susceptibility, which determines the reflection and
transmission coefficients for a single QW, is replaced with its
value averaged over the distribution of the exciton frequencies
along the plane of a QW:
\begin{equation}\label{eq:suscs_defs}
  \chi_{h,d}(\omega) = \int d\omega_0 f_{h,d}(\omega_0)
   \frac{\Gamma_0}{\omega_0 - \omega -
  i\gamma}.
\end{equation}
Here $\Gamma_0$ is the effective radiative rate of a single QW,
characterizing the strength of  the coupling between excitons and
electromagnetic field, $\gamma$ is the parameter of the
nonradiative homogeneous broadening, and $f_{h,d}$ are
distribution functions of the exciton energies of the host and
defect QW's respectively. The variance of this function, $\Delta$,
is interpreted as the parameter of the inhomogeneous broadening.
The functions $f_h$ and $f_d$ differ in their mean values, which
are $\omega_h$ and $\omega_d$ for the host and defect QW
respectively. The defect-induced effects are most pronounced if
$|\omega_h - \omega_d| \gg \Delta$. In this caseá the
inhomogeneous broadening of the host wells is negligible in the
vicinity of $\omega_d$, and defect-induced modifications of the
spectra can be studied with only  the inhomogeneous broadening of
the defect well taken into account.

If the length of the BMQW structure is not very large (for
$GaAs/AlGaAs $ it should be less than $\simeq 500$
periods\cite{ErementchoukPRB2003}), the reflection coefficient can
be presented in the form
\begin{equation}\label{eq:shallow_reflection}
r = \frac{i\bar\Gamma}{\omega_h - \omega + i(\gamma +
 \bar\Gamma)}\,
  \frac{\Omega_s -\Gamma_0 D_d}{i \Gamma_0 - \Gamma_0 D_d},
\end{equation}
where $D_d = 1/\chi_d$, $\bar\Gamma$ is the  radiative width of
the pure BMQW structure,\cite{IvchenkoMQW}
\begin{equation}\label{eq:shallow_Gamma_bar}
 \bar\Gamma = \Gamma_0 N/(1- i \pi q N),
\end{equation}
and $\Omega_s = (\omega_d -\omega_h)/N$.

\begin{figure}
  \includegraphics[width=3in, angle=-90]{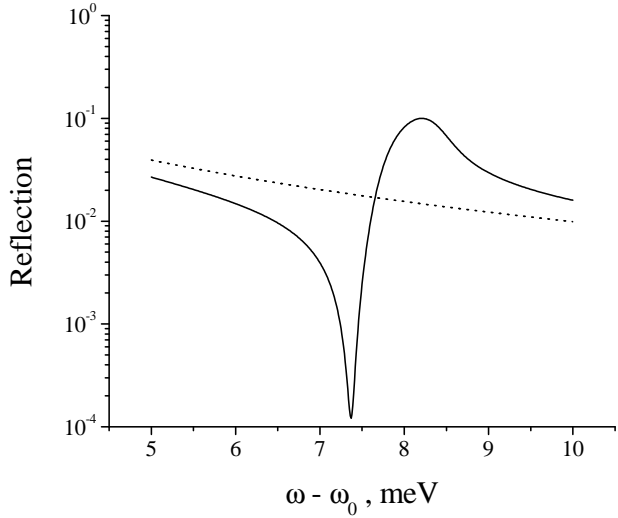}\\
  \caption{A typical dependence of the reflection coefficient on frequency
  is shown for the BMQW structure with $\Omega$-defect and length
  $N = 15$ (solid line). For reference, the reflection coefficient
  of a pure system is provided (dotted line).}\label{fig:shallow}
\end{figure}

The reflection spectrum is characterized by the presence of a
minimum and a maximum, Fig.~\ref{fig:shallow}, both of which lie
in the vicinity of $\omega_d$, but are shifted with respect to it.
The position of the minimum is determined mostly by parameter
$\Omega_s$, $\omega_{min} = \omega_d - \Omega_s
-\gamma^2/\Omega_s$. When $\Omega_s > \Delta$ the inhomogeneous
broadening of the defect well does not affect the value of the
reflection at $\omega_{min}$. For this to happen, the number of
wells must satisfy the condition $N<N_c$, where $N_c \approx 27$
for $GaAs/AlGaAs$,\cite{GaAs} and $N_c \approx 36$ for
$CdTe/Zn_{0.13}Cd_{0.87}Te$.\cite{CdTe} In this case, the value of
the reflection at the minimum $R_{min}$ is determined by the rate
of the non-radiative relaxation,
\begin{equation}\label{eq:shallow_reflection_minimum}
  R_{min} = |\bar\Gamma|^2\gamma^2 N^4/[(\omega_d -
  \omega_h)^4(N - 1)^2],
\end{equation}
and can be very small, when the latter is small. Meanwhile,
$\omega_{max}$ lies in the spectral region, where the host system
is almost transparent, and the value of the reflection at this
frequency, $R_{max}$ can be estimated as that of a single defect
QW
\begin{equation}\label{eq:shallow_reflection_alone}
  R_0 = 4\Gamma_0^2/(\pi\tilde \gamma + 2\Gamma_0)^2.
\end{equation}

\begin{figure}
  \includegraphics[width=2.8in,angle=-90]{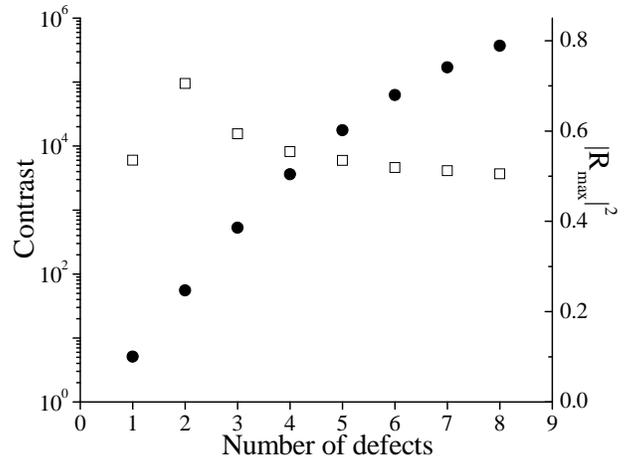}\\
  \caption{Dependencies of the maximal reflection (filled circles, right scale) and the
  contrast (empty squares, left scale) upon the number of the
  defects in BMQW structures.}\label{fig:maximum_multiple_defect}
\end{figure}

The  highest values of the contrast, defined as the ratio of the
maximum and minimum reflections $\eta = R_{max}/R_{min}$
\begin{equation}\label{eq:shallow_contrast}
  \eta \approx \left[({\omega_d - \omega_h})/{
  N\sqrt{\gamma \tilde \gamma}}\right]^4
\end{equation}
are obtained when the number of periods in the structure is small.
For low temperature values of $\gamma$, the contrast can be as
large as $10^{4}$. However, these large values of the contrast are
accompanied by rather small values of $R_{max}$. For switching or
modulating applications, it would be useful to have large
contrast, and a large maximum reflection. The latter can be
improved by considering structures with multiple defect wells.
This leads, of course, to an increase in the total number of
wells, but as we show, one can achieve a significant increase in
$R_{max}$ for quite reasonable total length of the structure
without compromising the contrast too much. Fig.~%
\ref{fig:maximum_multiple_defect} shows the results of numerical
computations of the dependence of  $R_{max}$ and the contrast upon
the number of defects. The structures were constructed of several
blocks, each of which is a $9$-period long BMQW with a single
defect well in the middle.

One can see that, indeed, the spectrum of such multi-defect
structures exhibits large $R_{max}$ (up to $0.8$ for structures no
longer than $80$ periods), while preserving high values of the
contrast (of the order of $10^{4}$).

\paragraph*{Tunability}

Applications of DBMQW structures for switching or modulating
devices is based on the possibility to change the value of the
reflection coefficient at a working frequency $\omega_w$ by
switching between $\omega_w = \omega_{max}$ and $\omega_w =
\omega_{min}$, using for instance the QCSE in order to change the
value of $\omega_d$. The structures under consideration also allow
for tuning of the working frequency of the device by shifting the
entire spectrum of the structure using QCSE in host wells. There
are several different ways to implement this idea, but here we
only want to demonstrate its feasibility. The main difficulty
results from the fact that shifting $\omega_h$ will detune the
whole system from the Bragg resonance and may destroy the
desirable spectral features discussed above. In order to see how
the detuning affects the spectrum, we assume for simplicity that
$\omega_h$ and $\omega_d$ change uniformly, and study the
reflection spectrum of an off-Bragg structure.

It was shown in Ref.~\onlinecite{DLPRBSpectrum} that the small
detuning from the Bragg resonance results in opening up a
propagating band at the center of the forbidden gap significantly
complicating the spectrum. It turns out, however, that as long as
$\omega_{min}$ and $\omega_{max}$ are well separated from
$\omega_h$, the detuning did not affect the part of the spectrum
associated with the defect.  Indeed, we show that the reflection
spectrum of an off-Bragg structure is described by the same
Eq.~(\ref{eq:shallow_reflection}) as that of the Bragg structure.
The only modification is the change of the definition of
$\bar\Gamma$, which now becomes
\begin{equation}\label{eq:off_effective_coupling}
  \bar \Gamma = \Gamma_0 N/[1 - i N
  \sin\pi(\omega-\omega_B)/\omega_B].
\end{equation}
Thus, for such shifts of the exciton frequencies, $\omega_s$, that
satisfy the condition
\begin{equation}\label{eq:off_condition}
  N \sin \left(\pi \omega_s/\omega_B \right)\ll 1
\end{equation}
the destructive effect of the detuning of the structure away from
the Bragg resonance is negligible in the vicinity of $\omega_d$.
It is important to note that the shift should be small in
comparison with the relatively big exciton frequency rather than,
for example, with the width of the reflection band. Because of
this circumstance, our structures can tolerate as large changes of
the exciton frequencies as are possible with QCSE. The result of
such a change is simply a  uniform shift of the part of the
spectrum shown in Fig.~(\ref{fig:shallow}) by $\omega_s$.

Additionally, Eq.~(\ref{eq:off_effective_coupling}) demonstrates a
stability of the considered spectrum with respect to weak
perturbations, such as small mismatch of refraction indices of
wells and barriers, different optical widths of the host and
defect quantum wells, and others.

\paragraph*{Conclusion}

In this paper we considered reflection spectra of one special case
of DBMQW structures, namely those in which the defect well differs
from the host wells in the value of the exciton energy. We showed
that if the frequency of the defect lies at the edge of the host
reflection band,  the spectrum of such a structure becomes
significantly modified: in the vicinity of the defect frequency it
becomes non monotonic with a well defined minimum and maximum. The
value of the reflection at the minimum, $R_{min}$ is determined
mostly by the rate of the non-radiative relaxation of excitons,
and can be very small at low temperatures. The small value of the
reflection leads to a giant contrast, defined as
$R_{max}/R_{min}$, which can be as large as $10^4$. The contrast
is one of the figures of merit for structures considered for
switching or modulating applications, however, the maximum
reflection $R_{max}$ in such structures is rather low. We showed
that $R_{max}$ can be significantly increased for structures with
several defects without compromising the value of the contrast.

An additional advantage of the proposed structures is their
tunability. We demonstrated that shifting the host and defect
exciton energies by several widths of the hosts' reflection band
leads to the uniform shift of the entire spectrum without any
significant adverse effects on the spectral region in the vicinity
of the defect frequency. This shift can be realized using, for
instance, QCSE so that the spectra of the considered structures
can be electrically tuned.

The work is supported by AFOSR grant F49620-02-1-0305 and PSC-CUNY
grants.

\end{document}